# Phase diagram and order-disorder transitions in $Y_{0.9}Gd_{0.1}Fe_2H_x$ hydrides ($x \geq 2.9$)

V. Paul-Boncour [a*], K. Provost [a], E. Alleno [a], A. N'Diaye [a], F. Couturas [a], E. Elkaim [b]

[a]Université Paris-Est Créteil, CNRS, ICMPE (UMR 7182), 2 rue Henri Dunant, 94320 Thiais, France
[b]Synchrotron SOLEIL, L'Orme des Merisiers, Saint-Aubin- BP 48, 91192 Gif-sur-Yvette, France

*Corresponding author, email: paulbon@icmpe.cnrs.fr, tel +33.1.49.78.12.07

**Abstract**

$Y_{0.9}Gd_{0.1}Fe_2$, which crystallize in a C15 cubic structure, can absorb up to 5 H/f.u. and its pressure-composition isotherm displays a multiplateau behavior related to the existence of several hydrides with different crystal structures. At room temperature $Y_{0.9}Gd_{0.1}Fe_2H_x$ hydrides ($2.9 \leq x \leq 5$) crystallize in three phases with cubic structure (C1, C2 and C3), two phases with monoclinic structures (M1 and M2), and one phase with orthorhombic structure (O), with the following sequence for increasing H concentration: C1, M1, C2, M2, C3, O. Each phase exists as single phase within a H homogeneity range, and they are separated from each other by two-phase domains. The reductions of crystal symmetry are related to various hydrogen orders into interstitial sites. Weak superstructure peaks were indexed by doubling the cubic cell parameter of the cubic C2 phase. Upon heating, the monoclinic M1 and M2 and the cubic C2 phases undergo order-disorder (O-D) transitions toward a disordered cubic structure $C_{Dis}$. These O-D transitions are reversible with thermal hysteresis effects. The cubic C3 and orthorhombic O phases transform into a disordered cubic phase accompanied by H desorption.

## 1. Introduction

Laves phases form many compounds which offer a broad range of functional and structural applications as reviewed in Stein et al. [1]. $RM_2$ Laves phases ($R$=Sc, Zr, Hf, Rare-earth, $M$= transition metal) can crystallize in three structure types: cubic $MgCu_2$ (*Fd-3m*), hexagonal $MgZn_2$ or $MgNi_2$ (*P6_3/mmc*). Hydrogen insertion in these phases yields not only to a cell volume enlargement, but also to a large variety of crystal structures with lower symmetry due to hydrogen ordering into interstitial sites [2]. $RFe_2$ intermetallic compounds have been studied for their giant magnetostrictive properties [3-5], but also for their ability to store large hydrogen content, i.e. up to 5 H/f.u. [6-9]. Interestingly, hydrogen insertion in $RFe_2$ compounds yields strong magnetic properties changes [9-16]. Depending on the hydrogenation conditions, $RFe_2$ hydrides can be either amorphous or crystalline [17-19]. In this work, we are interested in crystalline hydrides with a hydrogen induced lowering of their crystal symmetry.

Indeed, studies of the $YFe_2$ hydrides (H) and deuterides (D) revealed the existence of several phases with different structures related to that of the cubic $MgCu_2$ (C15) structure of the parent compound [7, 20-25]. Depending on the H(D) content, these structures are cubic, tetragonal, monoclinic or orthorhombic and display in some cases superstructures [23, 24, 26]. These phases exist for some well-defined H(D) content and are separated by two phase ranges. These distortions and/or superstructures disappear above a given temperature ($T_{O-D}$) to form a disordered cubic structure related to the loss of long range order of H or D atoms [27].

$YFe_2H_x$ hydrides were also studied for their magnetic and magnetocaloric properties especially around 4.2 H/f.u., where a ferromagnetic–antiferromagnetic (FM-AFM) transition with an itinerant electron metamagnetic (IEM) character has been observed [20, 21, 28, 29]. This magnetic transition is accompanied by a cell volume variation and is sensitive to any cell volume change induced by an applied external pressure [30, 31], D for H isotopic substitution [32, 33] or $R$ for Y substitution [34-37]. For example, the investigation of the monoclinic $Y_{0.9}Gd_{0.1}Fe_2(H_{1-z}D_z)_{4.2}$ compounds indicated that upon H for D substitution the FM-AFM transition temperature is shifted to higher temperature, whereas $T_{O-D}$ is shifted to lower value [38].

To clarify the origin of these structural transitions, we have investigated more systematically the influence of the hydrogen concentration on the structural properties of $Y_{0.9}Gd_{0.1}Fe_2H_x$ hydrides for $x \geq 2.9$ H/f.u.. For this purpose, several $Y_{0.9}Gd_{0.1}Fe_2H_x$ hydrides ($2.9 \leq x \leq 5$), and for comparison a $Y_{0.9}Gd_{0.1}Fe_2D_{4.2}$ deuteride, have been synthetized and characterized by combining pressure-composition isotherm, high resolution X-ray powder diffraction measurements using synchrotron radiation (SR-XRD) on the CRISTAL beam line of SOLEIL and Differential Scanning Calorimetry (DSC).

The room temperature phase diagram and the O-D structural transitions of $Y_{0.9}Gd_{0.1}Fe_2H_x$ hydrides and one deuteride are detailed in this work. The investigation of their magnetic and magnetocaloric properties will be the subject of a further study.

## 2. Experimental methods

$Y_{0.9}Gd_{0.1}Fe_2$ intermetallic compound have been prepared by induction melting of the chemical elements followed by three weeks annealing treatment under vacuum at 1100 K. The composition and homogeneity of the alloy was checked by X-ray powder diffraction (XRD) and electron probe microanalysis (EPMA) as described in [38].

$Y_{0.9}Gd_{0.1}Fe_2H_x$ and $Y_{0.9}Gd_{0.1}Fe_2D_{4.2}$ samples were prepared by solid-gas reaction using a Sievert apparatus as described in and pressure-composition isotherms (PCI) were measured. To reach H concentration above 4.2, a high-pressure device with maximum pressure of 0.8 GPa was used [39]. The samples were quenched into liquid nitrogen followed by a slow heating up to room temperature under air to poison the surface and prevent gas desorption.

The samples were characterized by XRD at room temperature with a Bruker D8 diffractometer. Synchrotron radiation measurements versus temperature were performed on the 2-circle diffractometer for selected samples using the CRISTAL beam line of SOLEIL (Saint Aubin, France). A multi-crystals (21) analyser was used to obtain high angular resolution diagrams and the wavelength, refined with a $LaB_6$ (NIST 660a) reference sample, was $\lambda = 0.582644$ Å. The powder samples were placed in sealed glass capillary tubes (0.3 mm diameter), which were rotated to ensure homogeneity. The samples were cooled or heated using a gas streamer cooler operating up to 380 K. All the SR-XRD patterns were refined with the Fullprof code [40].

The transitions were measured as a function of temperature using two differential scanning calorimeters (DSC) either a Q100 from TA instrument or a Perking Elmer Pyris Diamond instrument, with a rate of 10K/min.

## 3. Experimental results

### 3.1 Hydrogenation properties

The evolution of the absorption and desorption Pressure Composition Isotherm (PCI) of $Y_{0.9}Gd_{0.1}Fe_2H_x$ at 298 K is presented in Fig. 1. The presence of a plateau pressure is characteristic of the coexistence of two hydrides with different hydrogen concentrations, whereas the increase of pressure indicates a solid solution of hydrogen in one hydride phase. The present isotherm shows for pressure between 0.01 and 10 bars the existence of several plateaus with at least three different hydrides between 3.2 and 4.4 H/f.u.. The multiplateau isotherm of $RFe_2H_x$ compounds was correlated with the existence of phases with different structures [7, 22].

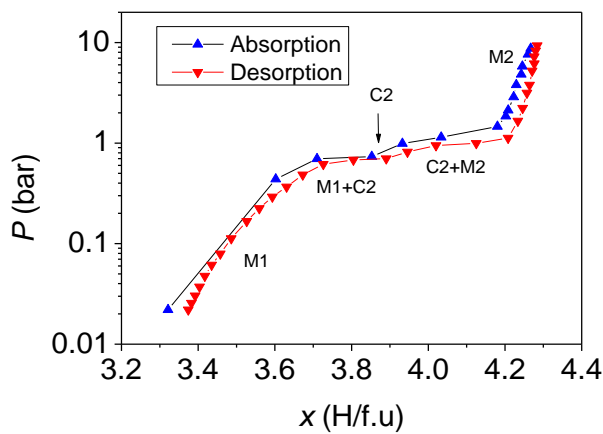

**Fig. 1.** Pressure-Composition isotherm of $Y_{0.9}Gd_{0.1}Fe_2H_x$ at 298 K. The phase abbreviations described in section 3.2 have been added.

### 3.2 Structural properties at room temperature

The analysis of the XRD data of several $Y_{0.9}Gd_{0.1}Fe_2H_x$ compounds at 300 K is summarised in Fig. 2, showing the evolution of the reduced cell volume per f.u. ($V/Z$) and the weight percentage versus H concentration for $2.9 \leq x < 5$. Six different hydrides, characterized by their crystal structure symmetry have been identified: three cubic hydrides (C1, C2, C3), two monoclinic hydrides (M1 and M2) and one orthorhombic hydride (O). The monoclinic and orthorhombic structures correspond to a distortion of the C15 cubic structure of the pristine compound. This distortion can be attributed to the hydrogen order into preferential interstitial sites, as previously observed for $YFe_2$ hydrides and deuterides by neutron powder diffraction (NPD) [23, 24, 33, 41].

The hydrides with cubic structures are intercalated between the monoclinic and orthorhombic hydrides. The space group and the cell parameters of the six phases described above are reported in Table 1. The cell parameters have been given for each phase at their maximum weight percentage. The three cubic XRD patterns were first refined with the same $Fd$-$3m$ space group (S.G.) than their parent compounds, with enlarged cell volumes. However, the close examination of the SR-XRD pattern of the C2 phase shows the existence of weak superstructure lines which can be indexed in a cubic cell with a double lattice parameter when, compared to $Y_{0.9}Gd_{0.1}Fe_2$. The two monoclinic patterns are refined with $C2/m$ S.G.; they are derived from the cubic structure as described for $YFe_2D_{3.5}$ [41] and $YFe_2D_{4.2}$ [24]. The orthorhombic hydride is described in the $Imm2$ space group and is isostructural to $YFe_2H_5$ and $ErFe_2D_5$ [9, 42]. Due to the presence of very high neutron absorbing Gd element and the large incoherent scattering of H it is difficult to perform neutron diffraction experiments on these $Y_{0.9}Gd_{0.1}Fe_2H_x$ hydrides. Although the localization of H(D) atoms cannot be obtained by XRD, the shift of the metallic atoms from their ideal position allows to observe the various structural transitions. As their XRD patterns are refined with the same structure types than non-substituted compounds, similar hydrogen ordering than for the $YFe_2$ deuterides can be assumed.

As presented in Fig. 2, the M1 hydride can be obtained as single phase between 3.2 and 3.5 H/f.u, whereas the other C2, M2 and C3 hydrides exist only for very narrow H concentration. All these phases are separated from each other by two phase domains. The three hydrides observed in the PCI curves correspond to M1, C2 and M2. The other phases are not identified in the PCI, as C1 absorb hydrogen below 0.01 bar whereas C3 and O synthesis require pressure much higher than 10 bars.

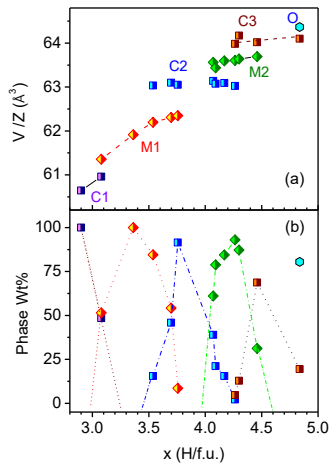

**Fig. 2.** (a) Cell volume per f.u. and (b) weight percentage of $Y_{0.9}Gd_{0.1}Fe_2H_x$ compounds at 300K.

**Table I**: Space group and cell parameters of the different $Y_{0.9}Gd_{0.1}Fe_2H_x$ hydrides at 300 K

| Phase | $x$ | S.G. | Z | $a$ (Å) | $b$ (Å) | $c$ (Å) | $\beta$ (°) | V/Z (Å$^3$) |
|---|---|---|---|---|---|---|---|---|
| C1 | 2.9 | *Fd-3m* | 8 | 7.8576(2) | | | | 60.6441(2) |
| M1 | 3.35 | *C2/m* | 4 | 9.5265(2) | 5.6623(1) | 5.5188(1) | 123.804(9) | 61.842(2) |
| C2 | 3.9 | *P-43m* * | 16 | 15.9207(1) | | | | 63.104(1) |
| M2 | 4.2 | *C2/m* | 4 | 9.4621 1) | 5.7538(1) | 5.5300(1) | 122.411(5) | 63.543(1) |
| C3 | 4.3 | *Fd-3m* | 8 | 8.0042(2) | | | | 64.101(1) |
| O | 4.8 | *Pmn*2$_1$ | 4 | 5.4417(1) | 5.8516(2) | 8.0855(3) | | 64.365(1) |

\* Space group used for pattern matching refinement

### 3.3 Order-disorder transitions

The SR-XRD patterns of the six selected samples containing most of the different phases described previously have been measured versus temperature and refined to study the order-disorder (O-D) transitions. The sample name and the results of the SR-XRD pattern refinement (reduced cell volume and weight percentage of each phase) near 300 K are reported in Table II. The corresponding refined patterns of these 6 samples are displayed in supplementary materials (Fig S1). The analysis of the SR-XRD patterns allow to observe the structural transitions from ordered structures toward disordered cubic structure upon heating (cooling). DSC measurements were also performed on the same samples to determine the transition temperatures and the enthalpy of reactions upon heating and cooling.

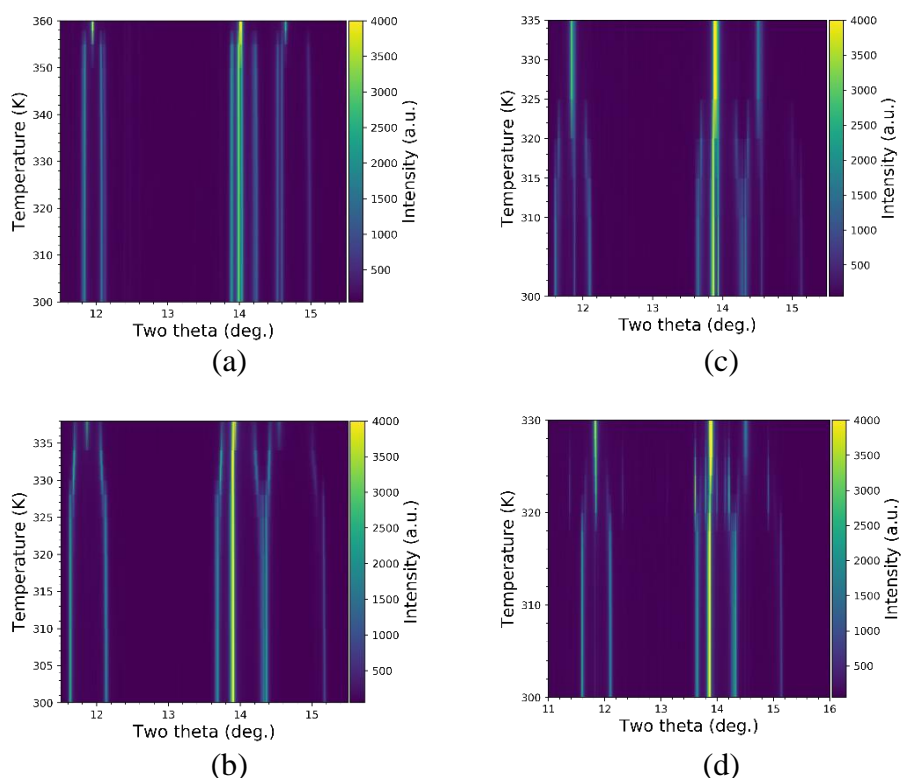

**Fig. 3.** Comparison of 2D plots of (a) HYD-M1, (b) DEUT-M2 (c) HYD-M2-A and (d) HYD-M2-B.

Different structural transitions have been found depending on the H concentration and the ordered structures. 2D plots of selected SR-XRD patterns upon heating are presented in Fig. 3 to illustrate the O-D transitions mainly for the monoclinic phases.

The evolution of the SR-XRD diffraction patterns versus temperature is reported in supplementary material for each sample (Fig. S2). In these plots the Bragg peaks are identified by their phase abbreviations. Above $T_{O-D}$ the patterns can be refined with one or two cubic phases which are call $C_{Dis}$ for disordered phases. The results are detailed below for each type of compound.

**Table II**: Reduced cell volume and weight percentage of each phase found in the SR-XRD patterns at 300 K of the $Y_{0.9}Gd_{0.1}Fe_2H(D)_x$ samples further studied as a function of temperature. The wt% of $Y_2O_3$ has been added for each compound.

| Sample name | $x$ (H/f.u.) | $T$ (K) | Phase | $V/Z$ (Å$^3$) | Phase Wt % | Wt% ($Y_2O_3$) |
|---|---|---|---|---|---|---|
| HYD-M1 | 3.35 | 300 | M1 | 61.841(2) | 98.5 | 1.5 |
| HYD-C2 | 3.9 | 300 | M1 | 62.349(1) | 8.1 | |
| | | | C2 | 63.052(1) | 91.9 | |
| DEUT-M2 | 4.2 | 300 | M2 | 63.087(1) | 98.5 | 1.5 |
| HYD-M2-A | 4.1 | 300 | C2 | 63.054(1) | 16.4 | 1.6 |
| | | | M2 | 63.559(1) | 57.2 | |
| | | | M'2 | 63.401(1) | 24.9 | |
| HYD-M2-B | 4.3 | 300 | C2 | 63.059(1) | 1.6 | 0.2 |
| | | | M2 | 63.610(1) | 90.7 | |
| | | | M'2 | 63.500(1) | 3.6 | |
| | | | C3 | 63.979(1) | 3.9 | |
| HYD-C3-O | 4.8 | 300 | C3 | 64.042(1) | 23.1 | 0.7 |
| | | | O | 64.317(1) | 76.2 | |

3.3.1 $Y_{0.9}Gd_{0.1}Fe_2H_{3.35}$

The $Y_{0.9}Gd_{0.1}Fe_2H_{3.35}$ hydride (HYD-M1) crystallizes in a single monoclinic structure up to 350 K. A transition from monoclinic to disordered cubic structure occurs between 350 and 360 K (Figs. 3*a* and 4). The existence of a narrow two-phase range and the existence of DSC peaks confirms the first order character of this O-D transition.

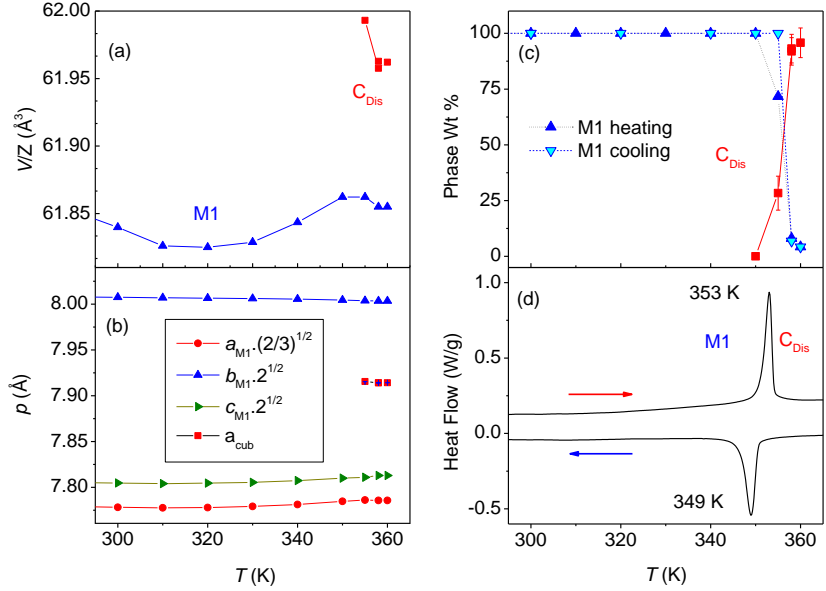

**Fig. 4.** (a) Cell volume (*V/Z*), (b) Cell parameters (p), (c) phase percentage and (d) DSC curve of $Y_{0.9}Gd_{0.1}Fe_2H_{3.35}$ (HYD-M1).

The lowering of crystal symmetry from cubic (*Fd-3m* S.G.) to monoclinic structure (*C2/m* S.G.) without distortion can be calculated according to:

$a_M = \sqrt{(3/2)} \cdot a_C$, $b_M = c_M = \sqrt{(1/2)} \cdot a_C$ and $\beta_M = 125.27°$ with $a_C = b_C = c_C = V^{1/3}$     (1)

To facilitate the comparison of the monoclinic cell parameters, they are plotted in Fig. 4 using an equivalent "cubic" description:

$a_M$ cubic $= \sqrt{2/3}\ a_M$ mono, $b_M$ cubic $= \sqrt{2} \cdot b_M$ mono, and $c_M$ cubic $= \sqrt{2} \cdot c_M$ mono     (2)

The refinement of the monoclinic cell parameters show a contraction of $a_{M1}$ (-1.6 %) and $c_{M1}$ (-1.3%), an elongation of the $b_{M1}$ parameter (+1.2 %) and a smaller monoclinic angle ($\beta_M = 123.8°$) compared to a non-distorted structure. A progressive reduction of the monoclinic distortion with a slight increase of $a_{M1}$, $c_{M1}$, $\beta_{M1}$ and a decrease of $b_{M1}$ is observed upon heating above 310 K. A jump of cell volume of 0.52 Å³ (+ 0.2 %) is seen between the monoclinic and the disordered cubic phases. The DSC curve shows one single peak at 353 K upon heating and 349 K upon cooling with enthalpy variations $\Delta H_{heat} = 9.2$ J/g and $\Delta H_{cool} = -8.2$ J/g respectively. A difference of 4 K of the maximum temperature is observed by DSC between heating and cooling indicating a small thermal hysteresis effect. These peaks can be attributed to the reversible M1-$C_{Dis}$ transformation which is observed at the same temperature upon heating.

3.3.2 $Y_{0.9}Gd_{0.1}Fe_2H_{3.9}$

The $Y_{0.9}Gd_{0.1}Fe_2H_{3.9}$ hydride (HYD-C2) remains cubic from 140 K up to 320 K with narrow linewidths, indicating a well crystallized compound (Fig. 5a). Weak superstructure peaks were observed and indexed in a primitive cubic cell with a doubling of the cell parameter. A fit using pattern matching mode of the pattern measured at 140 K (S.G. *P-43m* $a = 15.9207(1)$ Å) is presented in Fig. 5a, few peaks at low angle were also attributed to $Y_2O_3$. Due to the small intensity of the superstructure peaks and the large number of Wyckoff sites generated by the lowering of symmetry (9 Y, 10 Fe) and the number of atomic positions to refine (30) it was not possible to obtain a quantitative refinement in this cell. The superstructure peak intensities decrease upon heating and weak additional lines belonging to the M1 structure appears at 300 K. This transformation is reversible as the sample was first measured at 300 K, then cooled down to 120 K and heated again to 310 K. This evolution can be explained by a change of the limit of the M1-C2 hydrogen concentration upon heating. The DSC curve (Fig. 5b) shows peaks with maxima at 313 K upon first heating upon cooling and 307 K upon the second heating cycle indicating a first order reversible transformation. The variations of enthalpy ($\Delta H_{heat}$ = 2.7 J/g, $\Delta H_{cool}$ = 3.6 J/g) are smaller than for the M1-$C_{Dis}$ O-D transition.

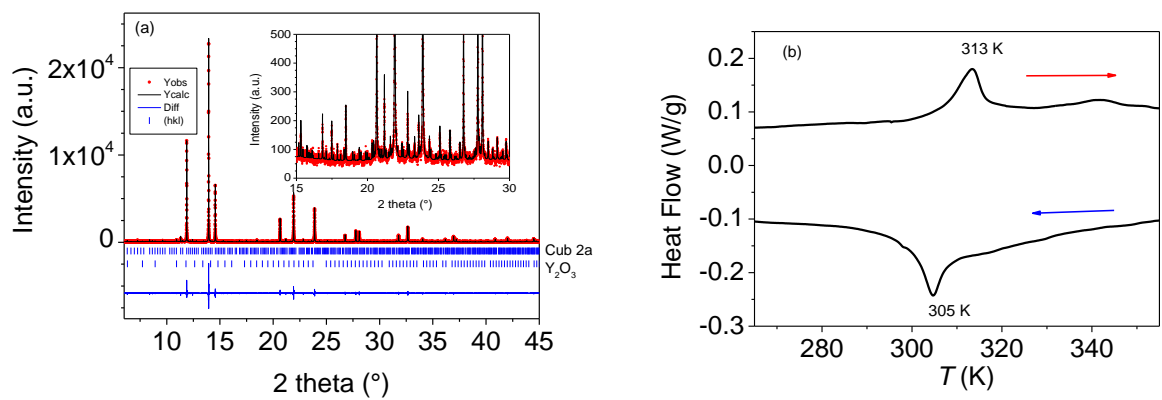

**Fig. 5.** (a) Pattern matching refinement of the $Y_{0.9}Gd_{0.1}Fe_2H_{3.9}$ pattern measured at 140 K using *P*-43*m* SG, (b) DSC curve.

3.3.3 $Y_{0.9}Gd_{0.1}Fe_2H_{4.2}$ hydrides and $Y_{0.9}Gd_{0.1}Fe_2D_{4.2}$

Several hydrides with a majority of the M2 phase have been synthetized. Contrary to the deuteride $Y_{0.9}Gd_{0.1}Fe_2D_{4.2}$ (DEUT-M2), and despite many trials, it was not possible to obtain 100 % of M2 phase for the $Y_{0.9}Gd_{0.1}Fe_2H_{4.2}$ hydride, as small fractions of the neighboring C2 and/or C3 phases were always present. This is explained by the very small plateau pressure difference between these phases in the $Y_{0.9}Gd_{0.1}Fe_2$-$H_2$ isotherm (Fig. 1). We have selected two samples: HYD-M2-A and HYD-M2-B containing 78 % and 90.7 % of monoclinic M2 phase at 300 K respectively and compared their evolution with that of $Y_{0.9}Gd_{0.1}Fe_2D_{4.2}$ as reference compound. It was also done to verify the isotope effect previously observed with a low resolution laboratory XRD device [38]. The evolution of the SR-XRD patterns of DEUT-M2, HYD-M2-A and HYD-M2-B upon heating is compared on the 2D plots in figs. 3b, 3c and 3d. Their reduced cell volume (*V/Z*), phase percentage obtained by Rietveld analysis and DSC curves are presented in Fig. 6, whereas their cell parameters are reported in Fig. 7.

For the deuteride the transformation from monoclinic M2 structure towards a cubic disordered phase $C_{Dis}$ occurs between 310 and 340 K with an intermediate monoclinic phase denoted M'2. The peaks belonging to the disordered cubic phase are broader than the monoclinic ones indicating probably a distribution of the hydrogen content inside the sample. The cell volume of the M'2 phase is slightly smaller than that of the M2 phase, whereas that of $C_{Dis}$ is larger (Fig. 6a).

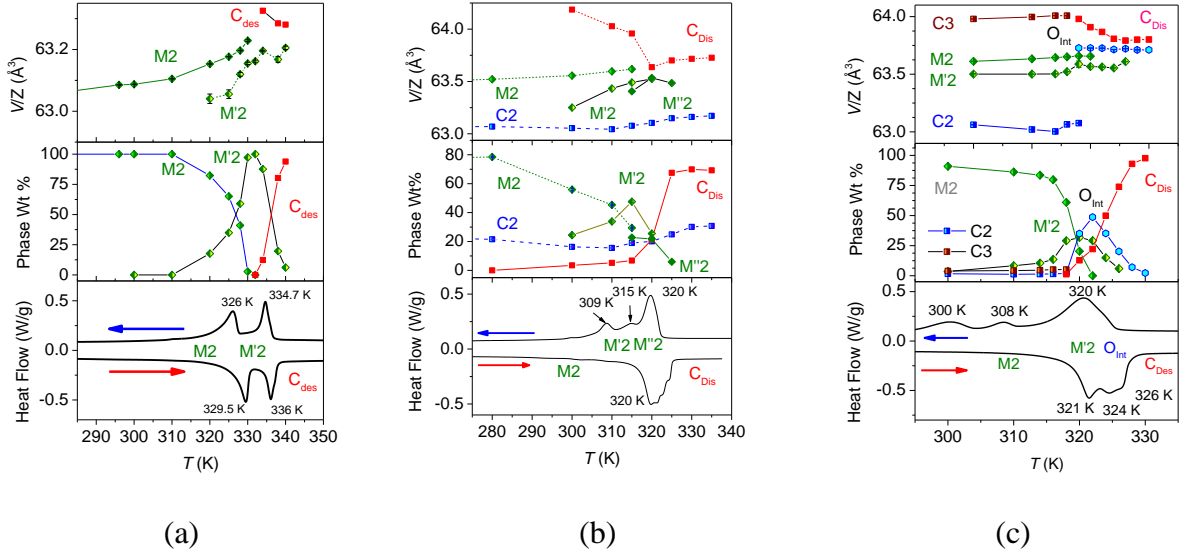

**Fig. 6.** Cell volume, weight percentage and DSC curve of (a) $Y_{0.9}Gd_{0.1}Fe_2D_{4.2}$ (DEUT-M2) and $Y_{0.9}Gd_{0.1}Fe_2H_{4.2}$ (b) (HYD-M2-A) and (c) (HYD-M2-B) compounds.

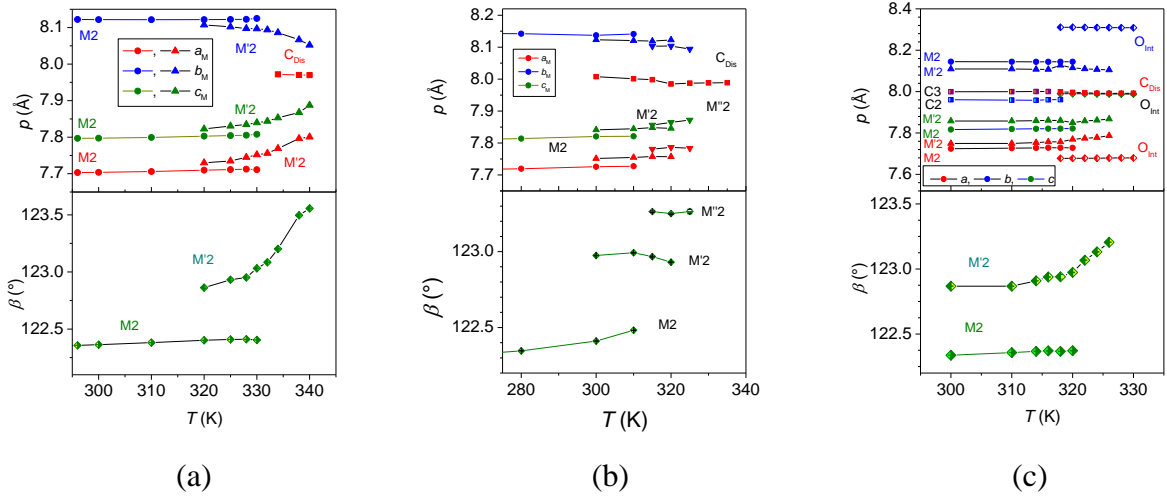

**Fig. 7**. Cell parameters ($p$) and monoclinic angle evolution of the M2, M'2, O and C2, C3 and $C_{Dis}$ phases in (a) DEUT-M2, (b) HYD-M2-A and (c) HYD-M2-B compounds.

The cell parameter variation, presented in Fig. 7 with equivalent cubic description as in Fig. 4 shows that the monoclinic distortion M2 induces a reduction of $a_{M2}$ (-3.2 %), $c_{M2}$ (-1.8 %) and $\beta_{M2}$ ($\beta$ = 122.4° or -2.3 %) parameters and an enlargement of $b_{M2}$ (+2%) compared to a non-distorted structure as was observed for the monoclinic M1 phase.

Upon heating $a_{M2}$, $c_{M2}$ and $\beta_{M2}$ parameters slightly increase whereas $b_{M2}$ decreases indicating a reduction of the distortion. M2 transforms into a second intermediate monoclinic phase M'2 between 320 K and 330 K with smaller monoclinic distortion. In addition, the monoclinic distortion of M'2 phase is more significantly reduced than that of M2 upon heating: $a_{M'2}$ (-2.9 to -2.1 %), $c_{M'2}$ (-1.7 to -1%) and $\beta_{M'2}$ ($\beta$ = 122.9° or -1.9 %) parameters and an enlargement of $b_{M'2}$ (1.9 to 1.1 %). At 332 K, peaks of a disordered cubic phase appear, and the compound is fully cubic above 340 K. It is important to notice that the monoclinic distortion of the M2 phase is larger than that of the M1 phase, which can explain that the O-D transition occurs through an intermediate M'2 phase to reduce the strains at the interphases. The DSC curve of DEUT-M2 is characterized by the presence of two peaks with maxima at 326 ($\Delta H_{heat}$ = 5.8(2) J/g) and 335 K ($\Delta H_{heat}$ = 5.3(2) J/g) upon heating and 329.5 ($\Delta H_{cool}$ = 5.7 (2) J/g) and 336 K ($\Delta H_{cool}$ = 5.0(2) J/g) upon cooling. These two peaks can be attributed to the M2-M'2 and then the M'2-$C_{Dis}$ transitions respectively, indicating their first order reversible character.

The sample HYD-M2-A shows a transformation from M2 towards the $C_{Dis}$ state through two intermediate M'2 and M"2 monoclinic phases between 300 and 325 K, i.e. at lower temperatures than for the deuteride (Fig. 6b). The XRD pattern analysis takes into account the presence of the cubic C2 phase which remains up to 335 K, and which total percentage and cell volume slightly increases above 310 K. As for the deuteride the $C_{Dis}$ cell volume is larger than the monoclinic one and decrease up to 320 K. The corresponding DSC curve in Fig. 6b (bottom) shows the existence of three less separated main peaks, which can be attributed to transitions from M2 to M'2, M'2 to M"2 towards $C_{Dis}$ occurring in the same relatively narrow temperature range. The cell parameter variation (Fig. 7b) shows the evolution of the monoclinic cell parameters. As previously, a progressive reduction of the monoclinic distortion is observed with an increase of $a_M$, $c_M$ and $\beta_M$ and a reduction of $c_M$. This occurs in a discontinuous way from M2 to M'2 and then through a second monoclinic phase M"2. It is difficult to estimate the variation of the enthalpy for each peak, as they are not clearly separated, a total value of $\Delta H_{heat}$ = 18.4(4) J/g and $\Delta H_{cool}$ = 17.9(4) J/g are obtained upon heating and cooling respectively. The DSC peaks are less separated upon heating than upon cooling, indicating a more important hysteresis effect than for the deuteride.

The HYD-M2-B sample at 300 K contains mainly the M2 phase and 1.6 wt% of C2 and 3.9 wt% of C3 phases (Fig. 7c). The analysis of the SR-XRD pattern shows a more complex structural evolution than for the two previous samples. This is the same hydride as in [38], but the better instrumental resolution of the synchrotron allow to separate more easily the different occurring phases and to perform a quantitative refinement of each phase. The transition from M2 to M'2 phase is observed between 295 K and 330 K, then the disordered cubic phase starts to appear above 320 K. In this sample an additional intermediate orthorhombic phase is observed between 317 and 330 K, as already observed in ref. [38]. It is obvious that this intermediate orthorhombic phase displays a significantly larger distortion than the monoclinic M2 and M'2) phases when compared in an equivalent cubic cell (Fig. 7c). Compared to the lattice parameters of the O phase obtained under high pressure, the parameters *a* (-0.24 %), *c* (-1.2 %) and *V* (-1.0 %) are smaller whereas its *b* parameter is slightly larger (+0.43 %). Its cell volume is intermediate between that of M2 and C3 phases, whereas its *c* parameter is close to the *a* parameter of C3 and $C_{Dis}$. This intermediate orthorhombic phase can occur either through a distortion of the C3 phase or because this intermediate orthorhombic phase exists in a limited temperature range between M2 and C3. Finally, a transformation of all distorted phases in one cubic phase $C_{Dis}$ is observed above 330 K. The transitions between all these phases is also observed by DSC (Fig. 7c, bottom), but in a more expanded temperature range upon cooling (with three clearly separated peaks between 290 and 330 K) than upon heating (Overlapped peaks between 304 and 333 K). The total enthalpy variations are $\Delta H_{heat}$ = 21.8(4) J/g and $\Delta H_{cool}$ = 21.9(4) J/g, i.e. slightly larger than for the previous monoclinic hydride.

3.3.4. $Y_{0.9}Gd_{0.1}Fe_2H_{4.8}$ hydride

The sample HYD-C3-O contains a mixture of cubic C3 and orthorhombic O phases at 300 K (Fig. S2 and Fig. 8). Upon heating a transformation from orthorhombic to disordered cubic structure with a line width broadening is observed. The shift and the broadening of the main cubic peak above 360 K indicates a partial hydrogen desorption, confirmed by the cell volume decrease (Fig. 8a). The transformation of $C_{Dis}$ in a mixture of M1 and C2 phases upon cooling confirms a significant hydrogen desorption (Fig. 8b). The cell parameters shows an expansion of *b* and *c* and a reduction of *a* parameters of the orthorhombic phase compared to the cubic parameter of the C3 phase (Fig. 8c). A small reduction of the orthorhombic distortion is observed before desorption. The DSC curve do not present any significant peaks at the transition. The intermediate orthorhombic phase $O_{int}$, observed in HYD-M2-B is not observed in the patterns. This confirms that $O_{int}$ exist in a small temperature range between M2 and C3, rather than to an intermediate phase between C3 and $C_{Dis}$.

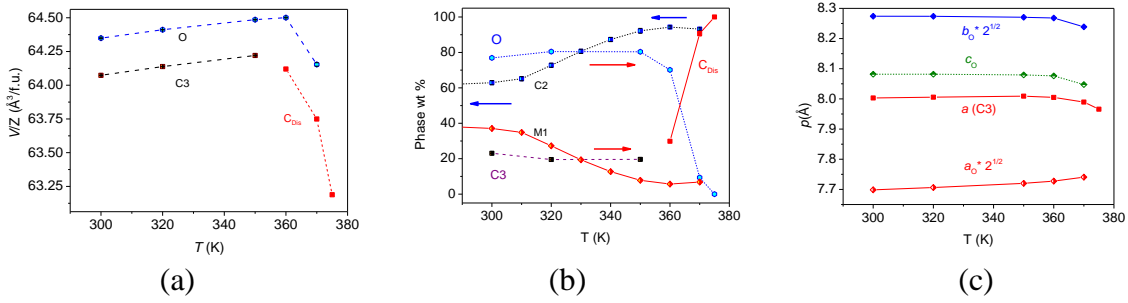

(a)            (b)            (c)

**Fig. 8.** a) Reduced cell volumes of C3, O and $C_{Dis}$ phases, b) weight percentage of C3, O, and $C_{Dis}$ phases upon heating and M1 and C2 upon cooling (c) cell parameters of C3 and O upon heating.

## 4. Discussion

The study of the structural properties of the $Y_{0.9}Gd_{0.9}Fe_2H_x$ hydrides for $x \geq 2.9$ allows to observe the existence of several phases with different structures derived from the pristine $Y_{0.9}Gd_{0.9}Fe_2$ compound below $T_{O-D}$. They should be related to the H(D) order in preferential interstitial $A_2Fe_2$ and $AB_3$ interstitial sites as previously observed for $YFe_2$ hydrides and deuterides.

Lowering of crystal symmetry upon H absorption have been observed for several Laves phases compounds [2]. The Laves phases contain many possible interstitial sites (96 $R_2M_2$, 32 $RM_3$ and 8 $M4$). Some geometric constrains such as a minimum radius of the interstitial site ($r \geq 0.4$ Å) [43] and a minimum H-H distance of 2.1 Å due to electronic repulsion between H atoms [44], exclude some sites. In addition, as the Fe-Fe distances in $YFe_2$ ($d_{Fe-Fe}$ =2.60 Å) are larger than the distances in the bcc Fe metal ($d_{Fe-Fe}$ =2.482 Å), the Fe atoms can be easily displaced from their ideal position [45]. The lowering of crystal symmetry can be related to the formation of more stable ordered compounds than disordered one at low temperature as observed through DFT calculations [38].

The Gd for Y partial substitution does not affect the formation of the two monoclinic phases M1 and M2, which were also observed for the non-substituted compounds. However, we have found that they are separated by an intermediate cubic phase C2. This C2 phase crystallizing with cubic superstructures inducing a doubling of the cell parameter is observed for the first time for $Y_{1-x}R_xFe_2$ compounds. A parallel can be made with the $YFe_2D_x$ phase diagram at low D concentration where the two tetragonal phases $YFe_2D_{1.2}$ and $YFe_2D_{1.9}$ are separated by a deuteride $YFe_2D_{1.75}$ also presenting a cubic superstructure with a doubling of the cell parameter [23]. Both Gd for Y and H for D substitution induces cell volume enlargement which probably stabilize this intermediate cubic phase. However, as the $YFe_2-H_2$ phase diagram was not studied with such details above 3.5 H/f.u. it would be interesting to investigate it again for comparison [22]. A second intermediate cubic phase C3 separates the monoclinic M2 and the orthorhombic phase O. However, no superstructures were observed for this C3 phase by SR-XRD.

Order-disorder transitions from low temperature ordered structures towards high temperature disordered cubic phase have been observed for the two monoclinic and the cubic C2 hydrides. These transitions present a first order character, as confirmed by DSC. The O-D transition of the M2 phase is more complex than for M1 and C2 as it occurs through different intermediate phases (M'2, M''2) presenting a progressive reduction of the monoclinic distortion. The appearance of an intermediate orthorhombic phase $O_{int}$ between the M2 and C3 phases, observed previously using laboratory XRD device for $Y_{0.9}Gd_{0.1}Fe_2(H_zD_{1-z})_{4.2}$ ($z$ = 0.75, 1) compounds [38] has been confirmed for HYD-M2-B sample. The origin of this intermediate orthorhombic phase reveals that a more complex phase diagram exists versus temperature and H content. Further in-situ measurements could be worth to precise its range of existence.

The range of O-D transition temperatures depends on the total H (D) content. The transition temperature are reduced as the H content increases as previously observed for $YFe_2$ hydrides and deuteride. $T_{O-D}$ were found equal to 460 and 444 K for $YFe_2D_{1.3}$ and $YFe_2D_{1.9}$ respectively [23], near 350 K for $YFe_2D_{3.5}$ [41] and between 330 and 340 K for $YFe_2D_{4.2}$ [24]. It is also sensitive to the isotope effect as the O-D transition occurs at a lower temperature for the M2 hydride (300-330 K) compared to the M2 deuteride (320-340 K). No O-D transition was observed for the C3 and O phases, either because they do not exist or because the H desorption occurs before the transformation. As the orthorhombic phase was synthetized at high pressure, it is a metastable phase stabilized by a quenching into liquid nitrogen to passivate the surface, this can explain the desorption occurring upon heating above 360 K.

Complex structural and magnetic phase diagrams have been also observed for $RMn_2$-$H_2$ systems [46-50]. But, the physical origin and the prediction of the various structural distortions depending of the $RM_2$ chemical composition and the hydrogen content remains still unsolved [2]. This would probably require detailed band structure calculations to compare the stability of each phase, but the complexity of the low temperature phase structure, with many different atomic positions and H interstitial sites makes it a difficult and time-consuming task. Nevertheless, this could be a very interesting subject to investigate in the future using new machine learning tools.

The magnetic and magnetocaloric properties of these compounds will be presented in a further study, with the goal to observe magnetocaloric effects near room temperature.

## 5. Conclusions

The $Y_{0.9}Gd_{0.9}Fe_2H_x$ hydrides (2.9 ≤ x ≤ 5) can crystallize in six different structures: C1, M1, C2, M2, C3 and O (C = cubic, M = monoclinic and O = orthorhombic). These compounds are formed for specific H content and are separated from each other by two phase ranges. The monoclinic M1, M2 and the cubic C2 undergoes reversible order-disorder transitions towards a disordered cubic structure. These O-D transitions are strongly dependent of the H content: the transition temperature decreases as the H content increases and is very sensitive to H(D) content for the M2 phase transformation with intermediate monoclinic phases. The C3 and O phases do not present O-D transition but start to desorb at 360 K.

**Acknowledgments**

We are thankful to the synchrotron SOLEIL for the allocated beam time on the CRISTAL beam line (proposal 20160868). We thank V. Shtender for his help to prepare the 2D plots presented in this work, L. Michely and B. Villeroy for their technical support for DSC measurements. We acknowledge also K. Hakkaoui for some sample preparation and C. Rakotoarimanana for her participation to the synchrotron experiments during their trainship.